\documentstyle[twoside, epsf]{article}

\input ibvs2.sty

\begin{document}
\IBVShead{5768}{04 May 2007}

\IBVStitle{13 New Eclipsing Binaries with Additional}
\IBVStitle{Variability in the ASAS Catalogue}

\IBVSauth{Pilecki, B.; Szczygie{\L}, D.M.} 

\IBVSinsto{Obserwatorium Astronomiczne Uniwersytetu Warszawskiego, Al.Ujazdowskie 4, 00-478 Warszawa, Poland \\
e-mail:pilecki@astrouw.edu.pl, dszczyg@astrouw.edu.pl}

\vskip 1cm

\textbf{Abstract.}
We present 13 new ASAS eclipsing binaries that exhibit additional periodic
variability due to pulsations, eclipses with another period or spots.
All contact and semi-detached binaries from the ASAS Catalogue were
investigated.

\vskip 1cm

\begintext

The All Sky Automated Survey has already collected over 6 years of observations for the majority 
of the sky (declinations $<$ +28 deg), down to 14th magnitude. Semi-automatic classification of  
variable stars resulted in the ASAS Catalogue of Variable Stars - ACVS (Pojma\'nski et al. 2006). 
For details on the classification procedure see Pojma\'nski (2002). A big part of ACVS consists of
eclipsing binaries, among them are 5,384 contact (EC), 2,957 semidetached (ESD), and 
2,758 detached (ED) binaries.
Recently a sub-sample of these has been searched for period changes (Pilecki et al. 2007).
During this investigation a side analysis was performed which resulted in 16 (13 new) binaries which are
suspect to additional 
periodic behaviour of various origin; secondary variability may be due to spots, pulsations, 
or second eclipsing binary in the system.
Two of them, namely 115143-6253.2 and 164802-6715.2, were found by D. Fabrycky, who pointed out
(private comm.) that these stars showed eclipses with another period.

\vskip 2mm

The search for second periodicity was performed on residual light curves of all EC and ESD 
binaries in ACVS (8,341 objects). After detecting an additional frequency for each object, all the 
light curves were sorted by amplitude of the frequency and the ones with a significant signal 
strength were inspected visually. This left us with 14 objects for which (together with additional
two stars mentioned above) a more detailed analysis was performed.

\vskip 2mm

In order to separate the light curves for both kinds of variability we applied an iterative method.
In the first step the best fitting model of an eclipsing binary ${\rm M_1}$ with orbital period 
${\rm P_1}$  was removed from the original light curve.
Then we analysed the residual light curve in the search for secondary period ${\rm P_2}$, which
was used to construct the model ${\rm M_2}$ of additional variability. This model was then
subtracted from the original light curve and the residual light curve was again investigated to find
a refined ${\rm M_1}$. After subtracting the new ${\rm M_1}$ from the raw ligt curve, the new 
${\rm M_2}$ was once again determined. In some cases one more step was performed to get a better 
model ${\rm M_1}$.

\vskip 2mm

Using residual light curves of models ${\rm M_1}$ and ${\rm M_2}$, variability was then classified
with periods ${\rm P_1}$ and ${\rm P_2}$ using the same procedure as in Pojma\'nski (2002). However,
all pulsating types were combined into one PULS category and, when it was plausible, we changed 
automatic classification to "Spot" type.

\vskip 0.8cm

\centerline{Table 1. ASAS eclipsing binaries exhibiting additional periodic variability.}
\vskip 3mm
\begin{footnotesize}
\begin{center}
\begin{tabular}{ccccccccc}
\hline
  ASAS ID      &$V_{max}$&${\rm P_{1}}$& Type& ${\rm P_{2}}$ & Type     &Blend& Other   & Other ID  \\
 (RA,DEC)      &  [mag]  &  [days]     &     &  [days]       &          & I A & data    &           \\
\hline
 174848-3503.5 &   7.45  & 7.71215     & ESD &     253.4     & PULS   & 0 0 & B3III     & V393 Sco    \\
 153713-1820.1 &   8.38  & 6.86170     & ESD &   6.87811     & Spot   & 0 0 & K1III,X   & IV Lib      \\
 103209-5905.7 &  10.50  & 0.953307    & ESD &  1.110270     & ESD=ED & 2 1 & F         & HD 302992   \\
 172738-3808.6 &  11.56  & 0.378603    & ESD & 0.423350     & EC/PULS & 2 2 & -----     &   -----     \\
 115143-6253.2 &   9.93  & 0.876114 & ESD    & 19.11(x2)     & ED     & 2 1 & B5        & BV 729      \\
 164802-6715.2 &  10.43  & 0.422509 & EC=ESD & 1.593378      & ED/ESD & 2 2 & -----     & TYC 9050-298-1  \\
 144001-1959.5 &  10.00  & 0.354445 & EC=ESD & 0.334349      & ESD/EC & 0 1 & G0,X      & BD-19 3931  \\
 031509-5144.2 &   9.61  & 21.4105  & EC/ESD & 21.1067       & Spot   & 1 0 & K1,X      & CD-52 646   \\
 125523-7322.2 &   9.74  & 206.1       & EC  & 250.2         &   ?    & 1 0 & -----     & TYC 9253-1392-1 \\
 103513-1206.5 &  11.43  & 0.384647    & EC  & 0.353901      & ESD/EC & 0 0 & -----     &   -----     \\
 131055-4844.0 &  10.80  & 7.06562     & EC? & 3.537421      & Spots? & 2 0 & ---,X     &   -----     \\
 103308-7133.8 &  10.58  & 0.816190    & EC  & 0.388607      & ESD=ED & 0 0 & -----     & TYC 9219-3329-1 \\
 190004-2741.4 &  12.24  & 0.439555    & EC  & 0.537903      & ESD/EC & 2 2 & -----     & V395 Sgr    \\
\hline
\end{tabular}
\end{center}
\end{footnotesize}

\vskip 2mm

In Table 1 we listed both periods (${\rm P_1}$ and ${\rm P_2}$), separate variability types and the
possible degree of blending (0 for none, 1 for small and 2 for large) listed in two columns, designated
by I and A. The first one (I) is the degree of blending evaluated subjectively by an examination of higher resolution images from Digitized Sky Survey, whereas A is the result of brightness comparison in different apertures of ASAS photometry. The radius of the smallest aperture is 1 pixel and for the largest 3 pixels, so two faint stars close to each other are separated when using small aperture and counted as one object when
using a large aperture, significantly increasing the brightness.
Some additional information from the SIMBAD database is given (if available) such as an other
identifier, spectral type, and whether the star might be an X-ray source (X).

Two stars were found in the WDS catalogue of astrometric doubles and multiples (Mason et al. 2001).
234520-3100.5 was identified as a double star (11.58 mag + 11.94 mag) with a separation of 1"
and 125523-7322.2 (10.6 mag + 11.5 mag) with a separation of 2.4".

\vskip 2mm

In the course of this analysis 7 out of 13 objects turned out to be double eclipsing binaries (ie.
quadruples that consists of two doubles), whereas one exhibits additional pulsations; for one object
both scenarios are probable.
There are also two stars whose secondary periods have values close to that of primary periods. This
kind of behaviour is believed to be due to spots on one of the binary's components. For the remaining two
we have no plausible explanation.

\vskip 0.8cm

\centerline{Table 2. Objects examined independently by Pigulski \& Michalska.}
\vskip 3mm
\begin{footnotesize}
\begin{center}
\begin{tabular}{ccccccccc}
\hline
  ASAS ID      & 2nd type &Blend& Other ID  \\
 (RA,DEC)      &          & I A &           \\
\hline
 182323-1240.9 &   PULS   & 2 0 & FR Sct    \\
 234520-3100.5 &  EC/PULS & 0 0 & -----     \\
 084350-4607.2 &  ESD/EC  & 2 2 & ALS 1135  \\
\hline
\end{tabular}
\end{center}
\end{footnotesize}

\IBVSfig{15cm}{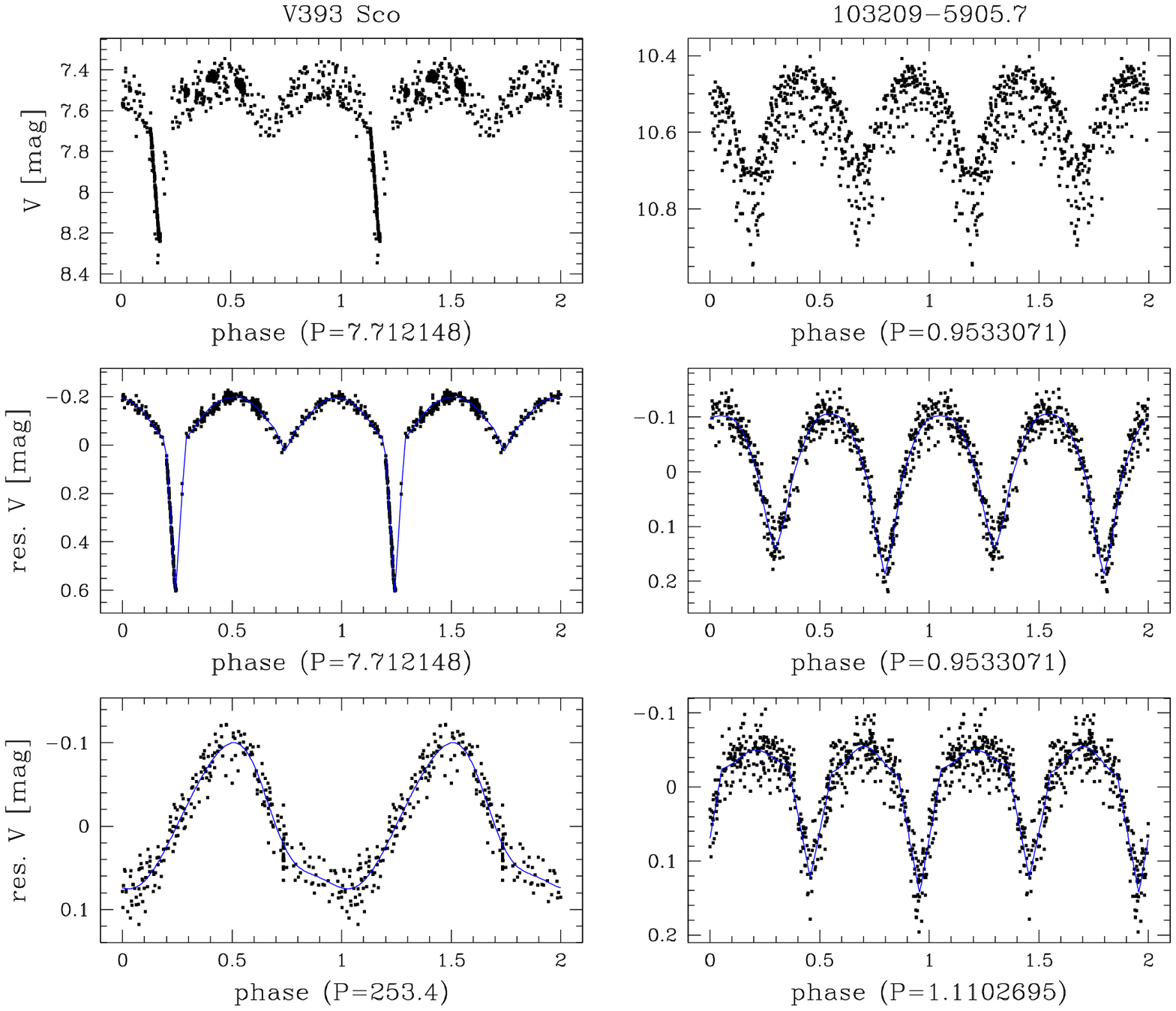}{
Two examples of double periodic behaviour. Original and residual light curves are shown.
The light curves of other stars are attached at the end of this paper in a simplified form.}

Three stars listed in Table 2 were independently found and recently analysed by Pigulski \& Michalska (2007a,b). They found FR Sct to be a triple VV Cephei-type system, 234520-3100.5 to show additional $\delta$ Scuti
behaviour, and 084350-4607.2 to exhibit $\beta$ Cephei-type variations. For them we quote only our second variability type and an estimation of a degree of blending.

\vskip 2mm

One star, namely 131055-4844.0, has a secondary period value close to (but not the 
same as) half the value of the primary variation period. Moreover, a residual light curve of the
second variability has an eclipsing-like shape with two minima of different depth. This cautions,
that the primary period may be two times smaller and the primary variability may be due to pulsations rather than eclipses.

\vskip 2mm

All presented stars need a further study. Spectroscopic and photometric observations of higher resolution
will help to determine a true nature of these objects.

\vskip 2mm

\textbf{Acknowledgements.} We would like to thank D. Fabrycky for pointing out two stars with
additional periodic behaviour. Our analysis made use of the Digitized Sky Survey images made available by the STScI. 
This work was supported by the MNiSW grant N203 007 31/1328.

\vskip 1cm

\references

Mason, B. D., et al. 2001, {\it AJ}, {\bf 122}, 3466

Pigulski, A., Michalska, G., 2007a, {\it IBVS}, 5757

Pigulski, A., Michalska, G., 2007b, {\it AcA}, {\bf 57}, 61

Pilecki, B., Fabrycky, D., Poleski, R., 2007, astro-ph/0703705, Accepted to MNRAS

Pojma\'nski, G.,  2002, {\it AcA}, {\bf 52}, 397

Pojma\'nski, G., Maciejewski, G., Pilecki, B., Szczygiel, D.,  2006, {\it VizieR On-line Data Catalog II/264.}

\endreferences

\IBVSfig{9cm}{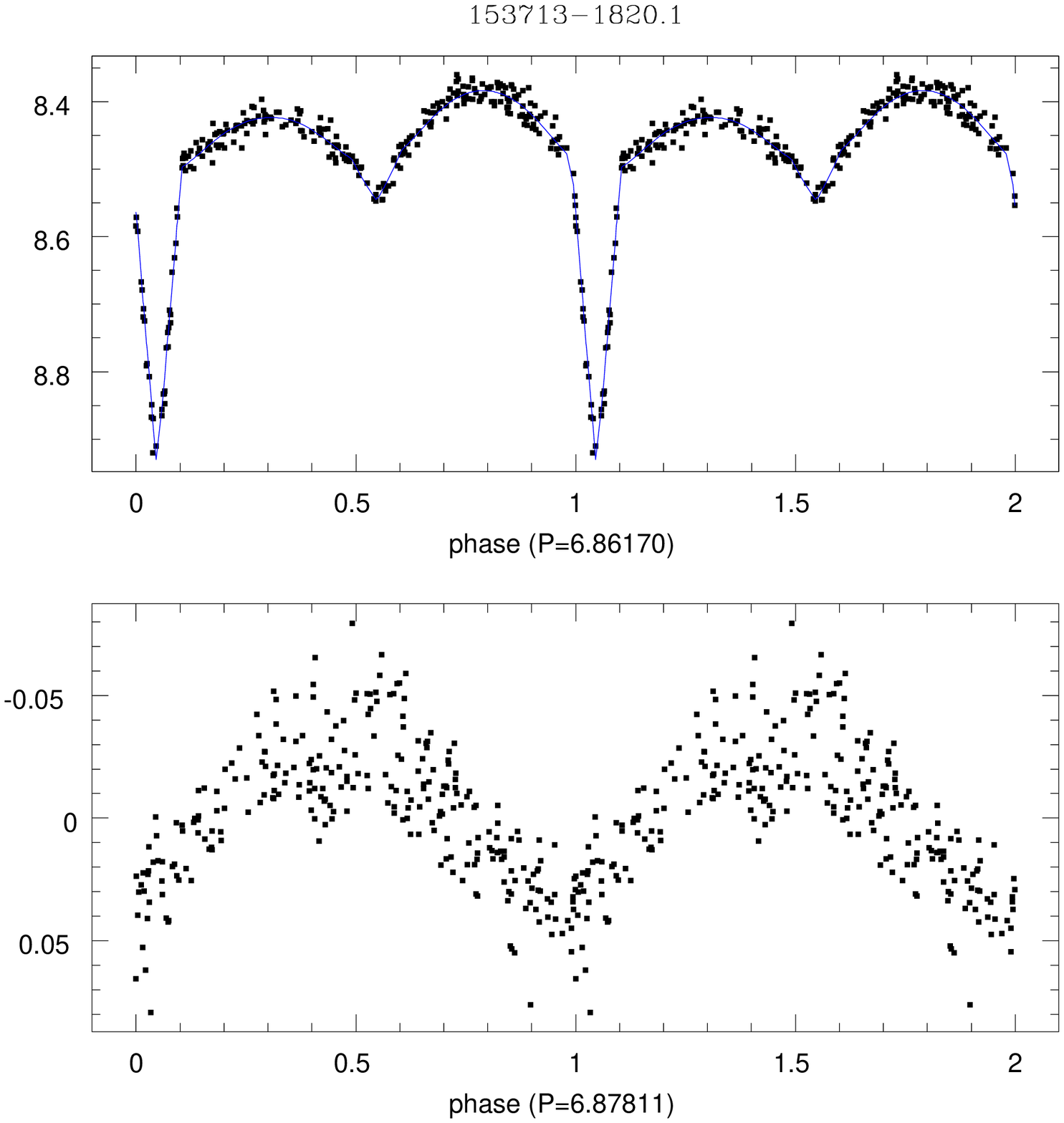}{Light curve of 153713-1820.1.}
\IBVSfig{9cm}{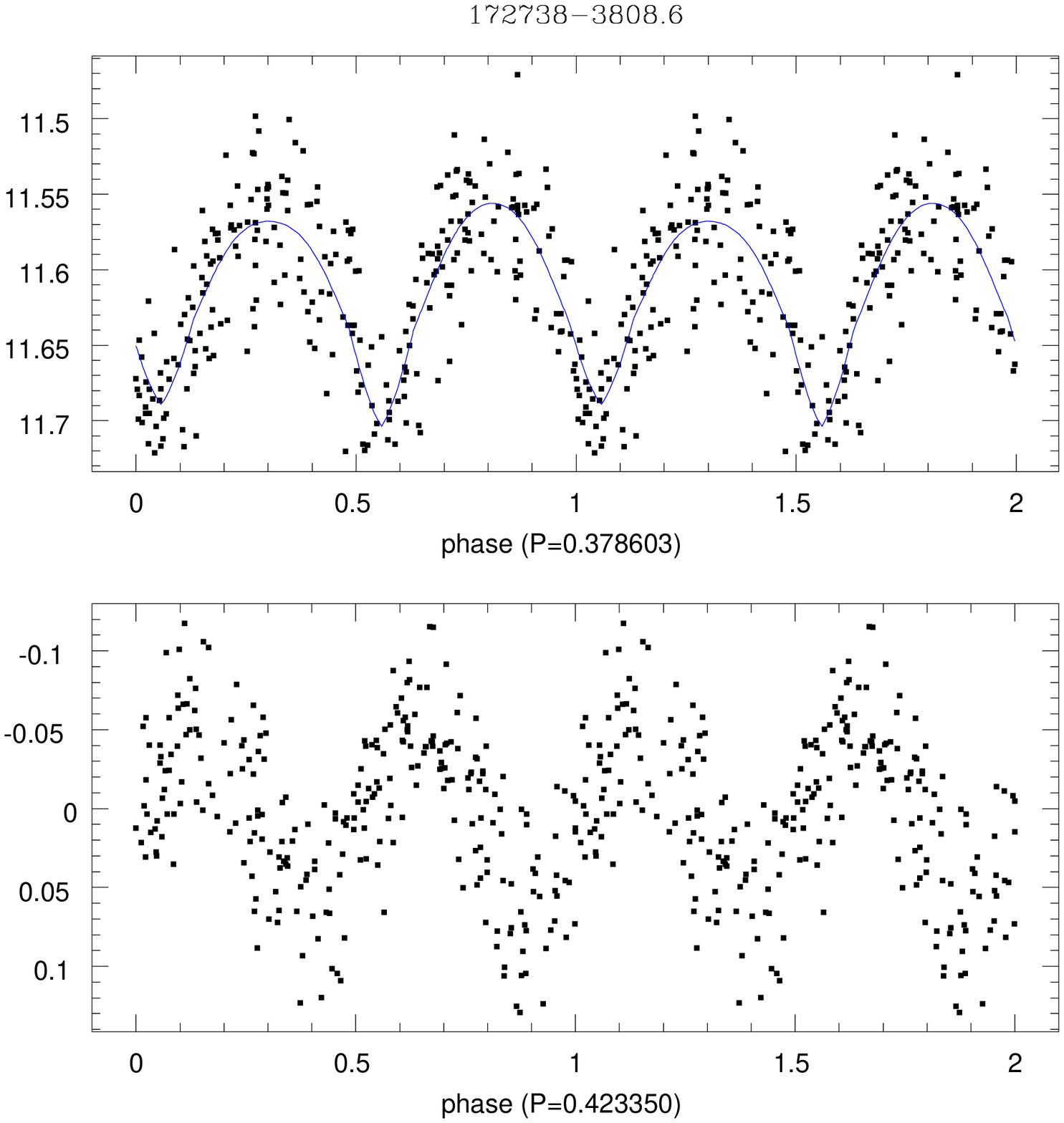}{Light curve of 172738-3808.6.}
\IBVSfig{9cm}{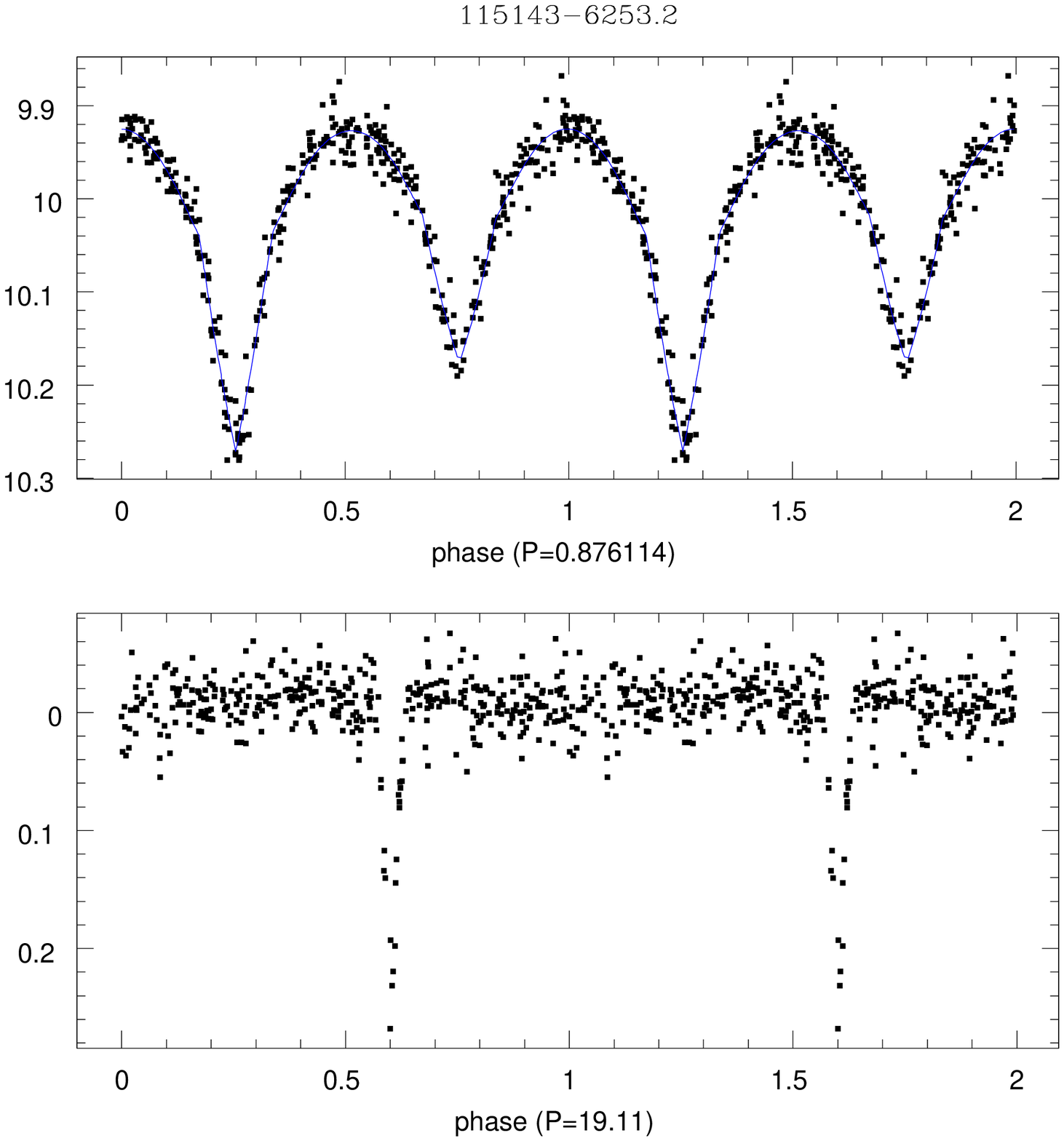}{Light curve of 115143-6253.2.}
\IBVSfig{9cm}{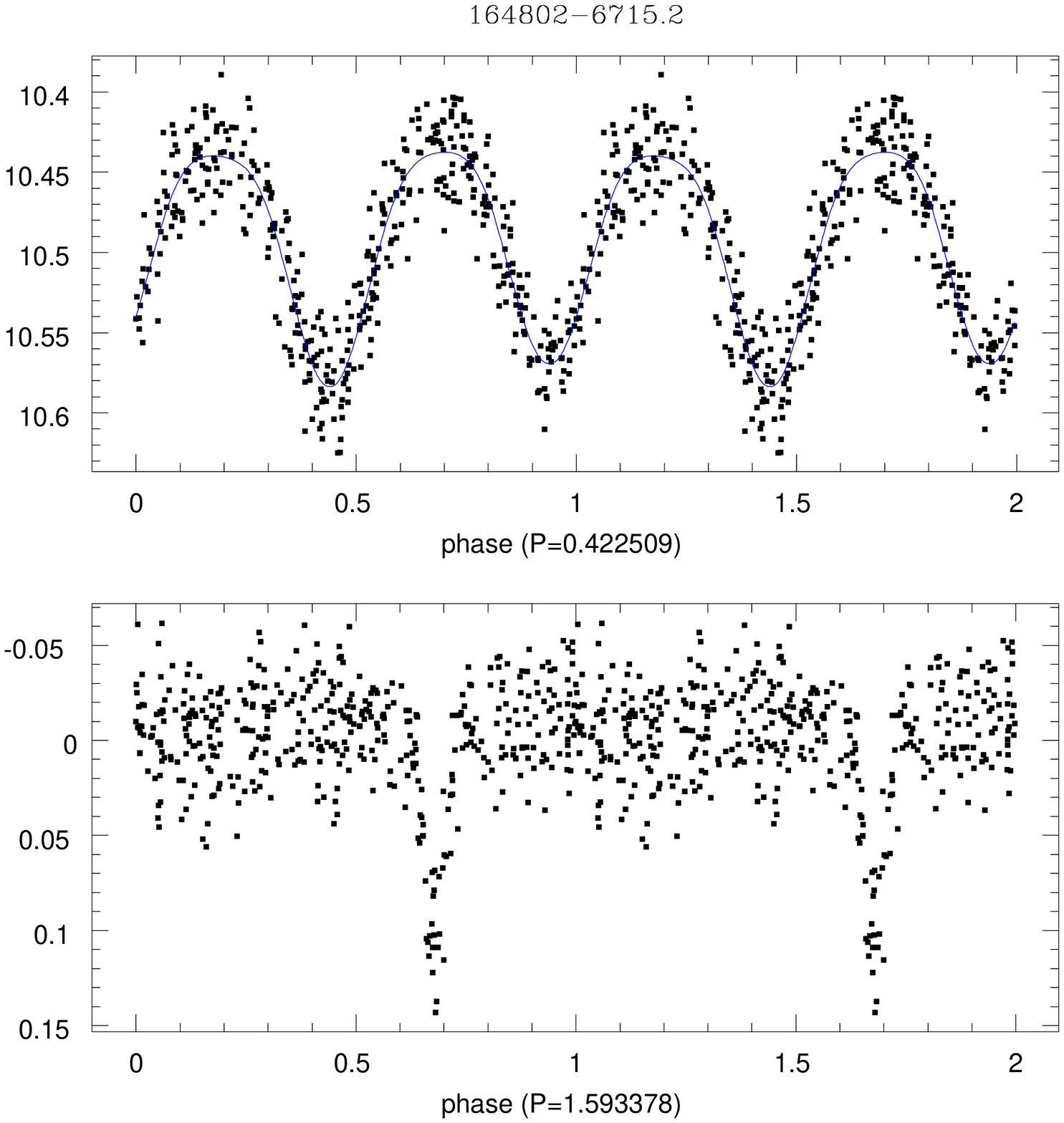}{Light curve of 164802-6715.2.}
\IBVSfig{9cm}{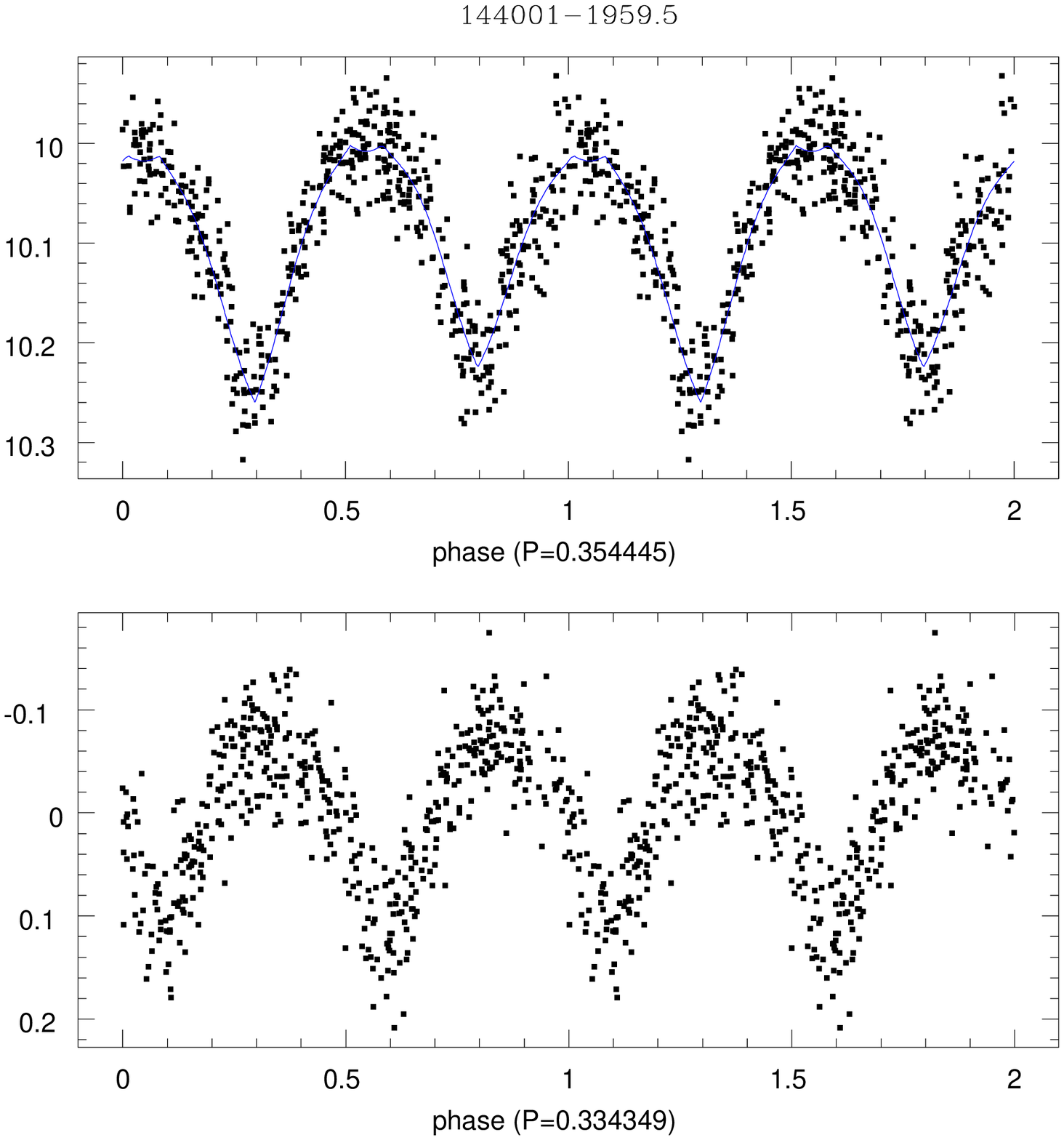}{Light curve of 144001-1959.5.}
\IBVSfig{9cm}{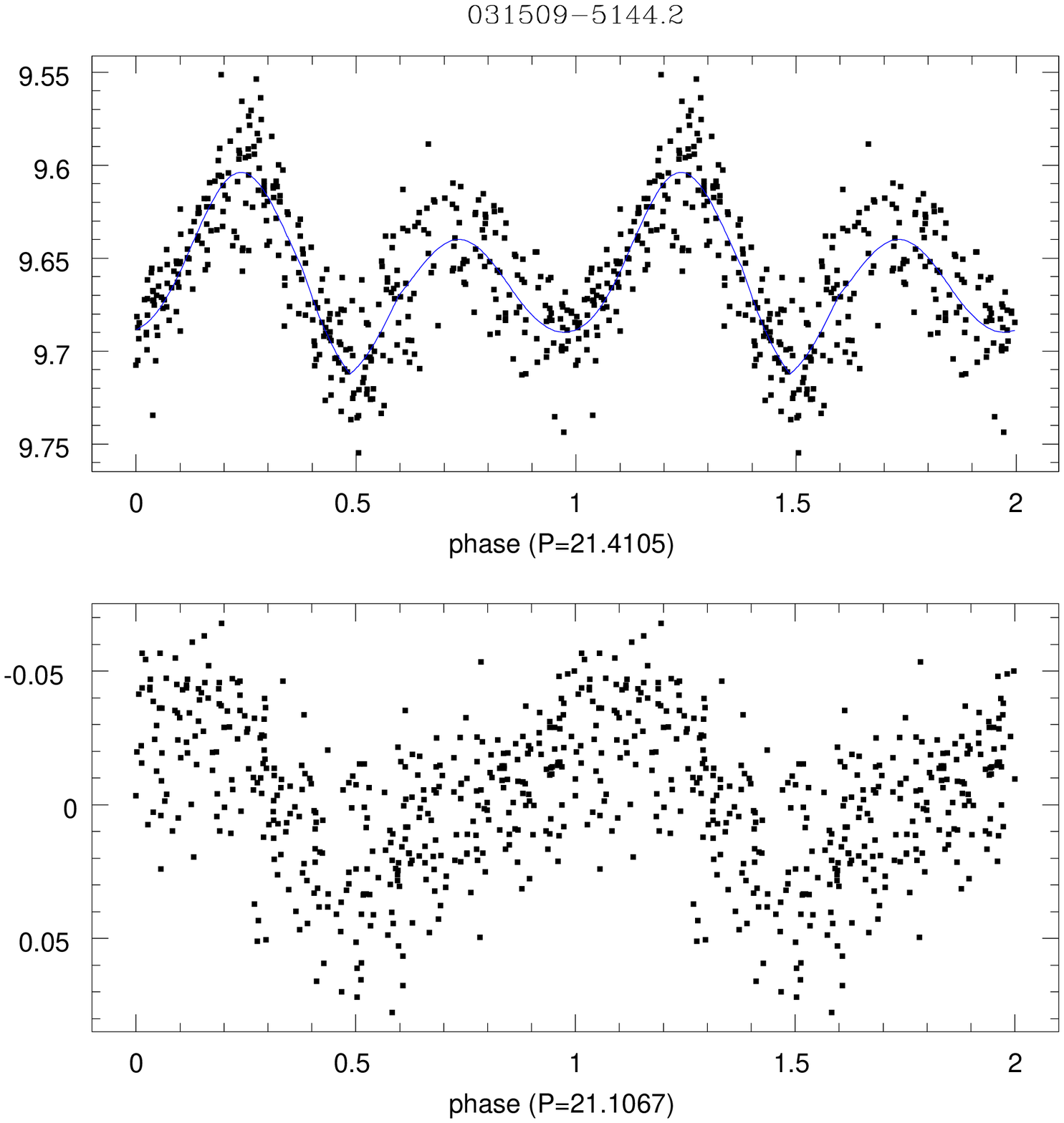}{Light curve of 031509-5144.2.}
\IBVSfig{9cm}{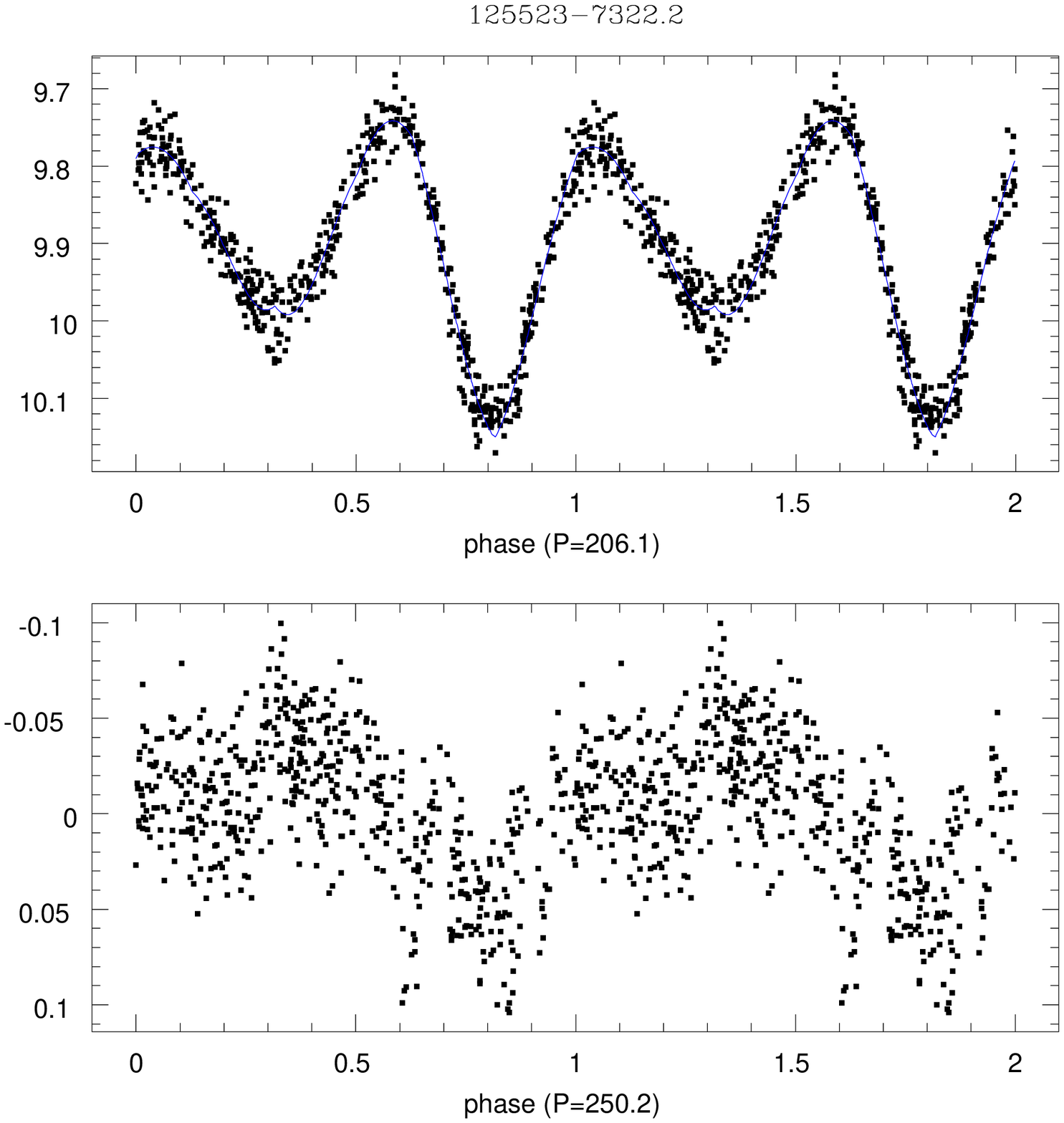}{Light curve of 125523-7322.2.}
\IBVSfig{9cm}{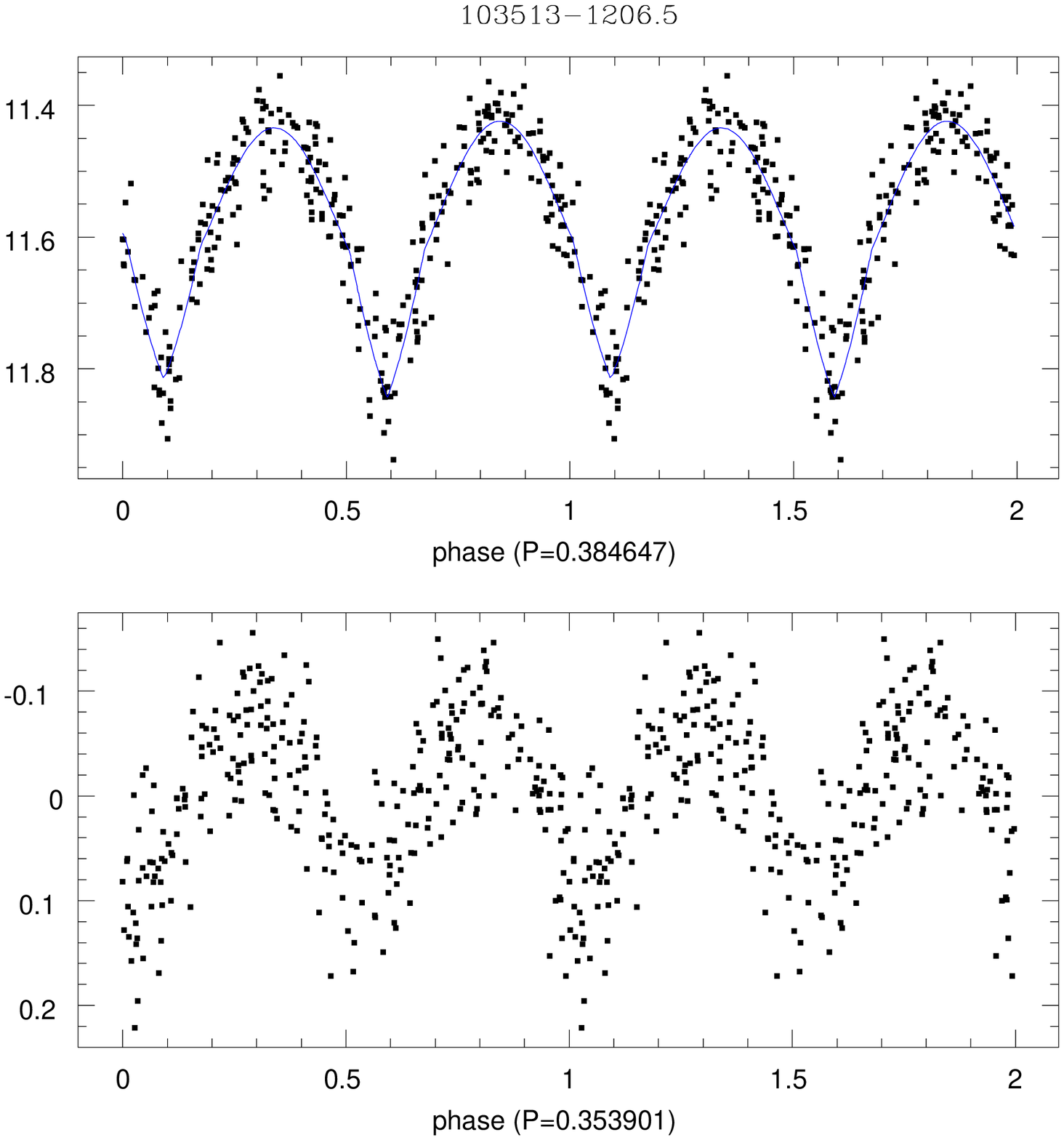}{Light curve of 103513-1206.5.}
\IBVSfig{9cm}{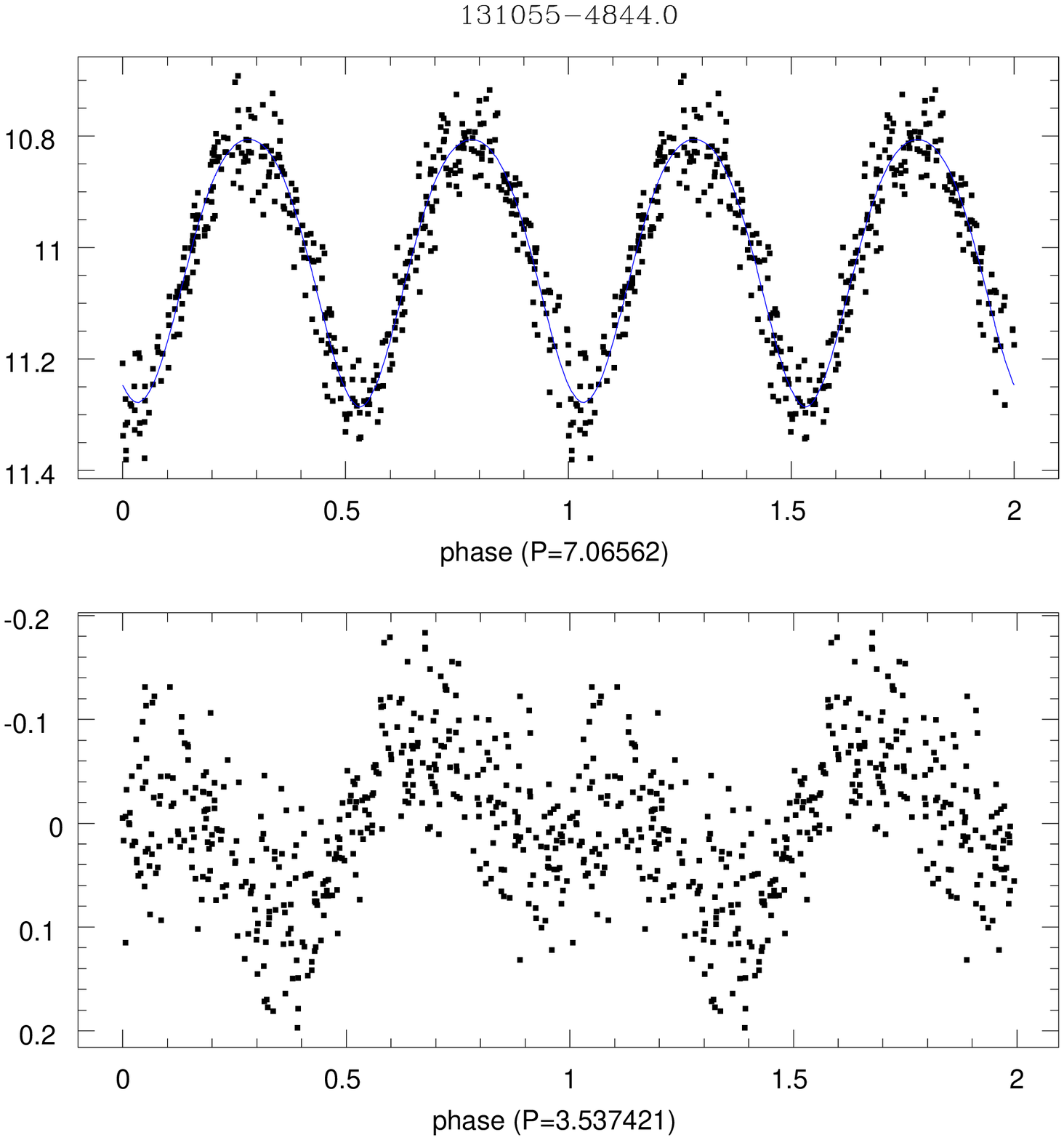}{Light curve of 131055-4844.0.}
\IBVSfig{9cm}{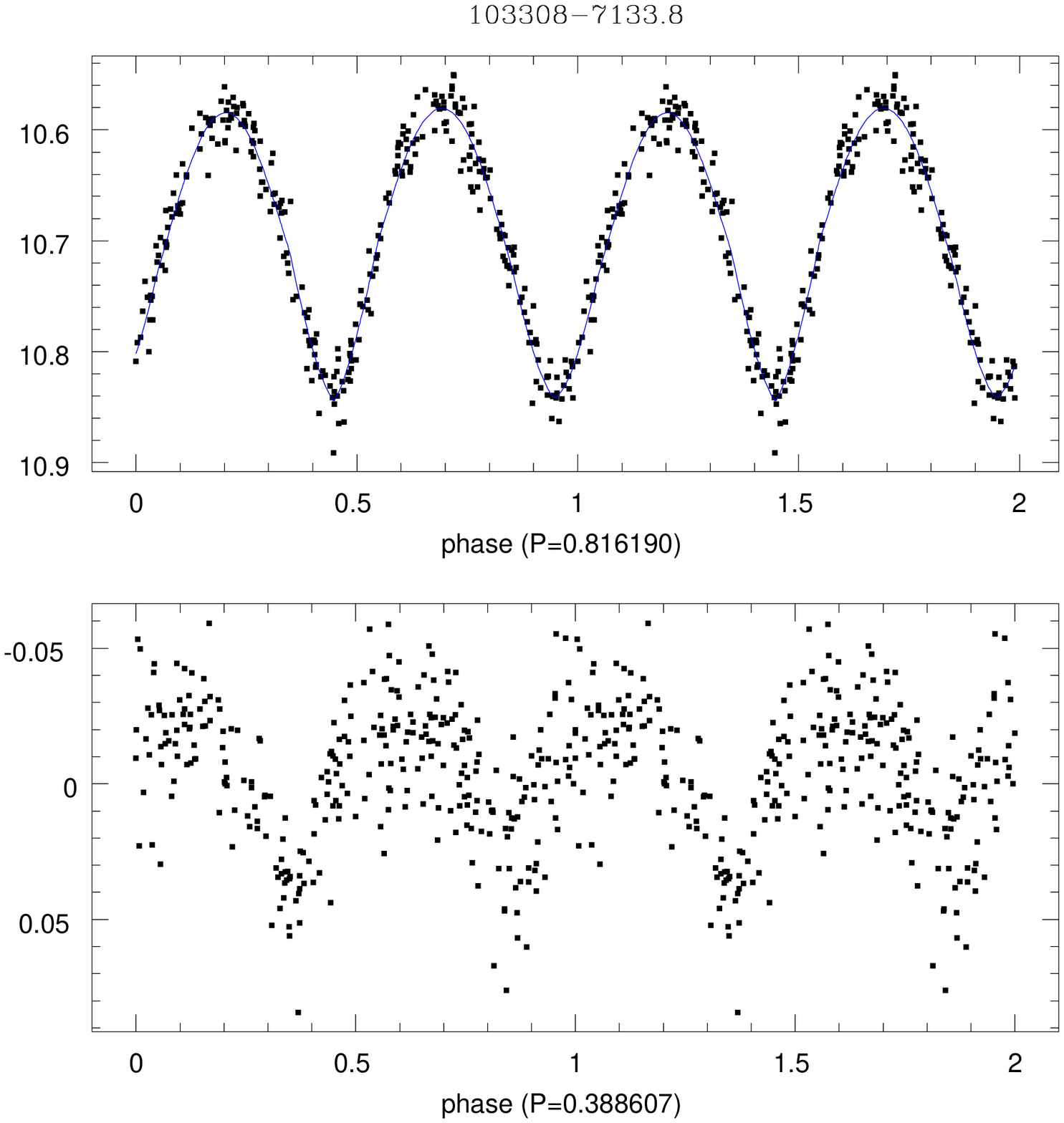}{Light curve of 103308-7133.8.}
\IBVSfig{9cm}{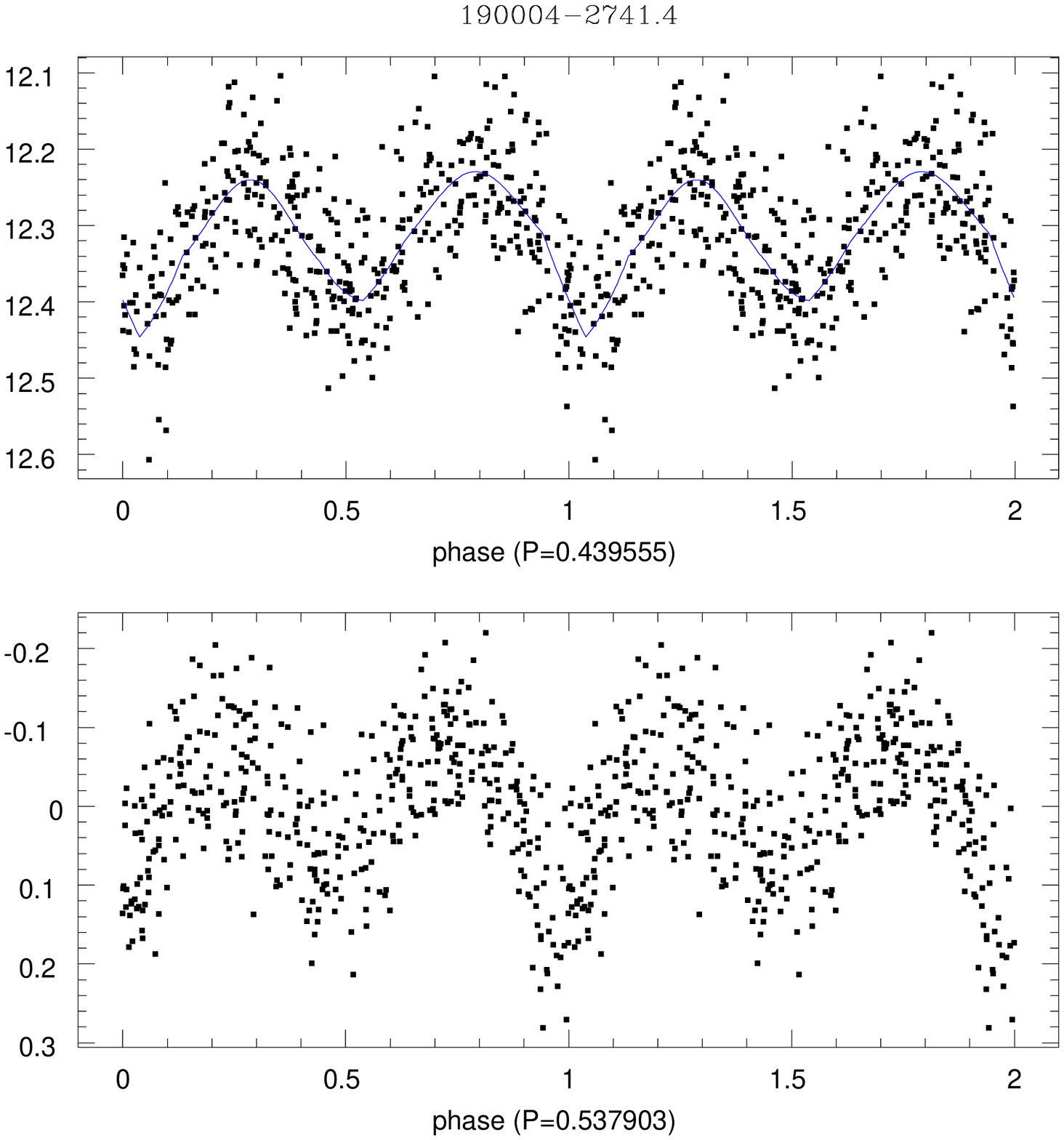}{Light curve of 190004-2741.4.}

\end{document}